# Transceiver Design with Low-Precision Analog-to-Digital Conversion : An Information-Theoretic Perspective

Jaspreet Singh, *Student Member, IEEE,* Onkar Dabeer, *Member, IEEE,* and Upamanyu Madhow, *Fellow, IEEE*

### Abstract

Modern communication receiver architectures center around digital signal processing (DSP), with the bulk of the receiver processing being performed on digital signals obtained after analog-to-digital conversion (ADC). In this paper, we explore Shannon-theoretic performance limits when ADC precision is drastically reduced, from typical values of 8-12 bits used in current communication transceivers, to 1-3 bits. The goal is to obtain insight on whether DSP-centric transceiver architectures are feasible as communication bandwidths scale up, recognizing that high-precision ADC at high sampling rates is either unavailable, or too costly or power-hungry. Specifically, we evaluate the communication limits imposed by low-precision ADC for the ideal real discrete-time Additive White Gaussian Noise (AWGN) channel, under an average power constraint on the input. For an ADC with $K$ quantization bins (i.e., a precision of $\log_2 K$ bits), we show that the Shannon capacity is achievable by a discrete input distribution with *at most* $K + 1$ mass points. For 2-bin (1-bit) symmetric ADC, this result is tightened to show that binary antipodal signaling is optimum for any signal-to-noise ratio (SNR). For multi-bit ADC, the capacity is computed numerically, and the results obtained are used to make the following encouraging observations regarding system design with low-precision ADC : (a) even at moderately high SNR of up to 20 dB, 2-3 bit quantization results in only 10-20% reduction of spectral efficiency, which is acceptable for large communication bandwidths, (b) standard equiprobable pulse amplitude modulation with ADC thresholds set to implement maximum likelihood hard decisions is asymptotically optimum at high SNR, and works well at low to moderate SNRs as well.

Jaspreet Singh and Upamanyu Madhow are with the Department of Electrical and Computer Engineering, University of California, Santa Barbara. Their research was supported by the National Science Foundation under grants ANI-0220118 and ECS-0636621, and by the Office of Naval Research under grant N00014-06-1-0066. E-mail: {jsingh, madhow}@ece.ucsb.edu

Onkar Dabeer is with the School of Technology and Computer Science, Tata Institute of Fundamental Research, Mumbai, India. His research was supported by Grant SR/FTP/ETA-16/2006 from the Department of Science and Technology, Government of India, and in part by the Homi Bhabha Fellowship. E-mail: onkar@tcs.tifr.res.in

## I. INTRODUCTION

Digital signal processing (DSP) forms the core of modern digital communication receiver implementations, with the analog baseband signal being converted to digital form using Analog-to-Digital Converters (ADCs) which typically have 8-12 bits of precision. Operations such as synchronization, equalization and demodulation are then performed in the digital domain, greatly enhancing the flexibility available to the designer. The continuing exponential advances in digital electronics, often summarized by Moore's "law" [1], imply that integrated circuit implementations of such DSP-centric architectures can be expected to continue scaling up in speed and down in cost. However, as the bandwidth of a communication system increases, accurate conversion of the analog received signal into digital form requires high-precision, high-speed ADC, which is costly and power-hungry [2]. One possible approach for designing such high-speed systems is to drastically reduce the number of bits of ADC precision (e.g., to 1-3 bits) as sampling rates scale up. Such a design choice has significant implications for all aspects of receiver design, including carrier and timing synchronization, equalization, demodulation and decoding. However, before embarking on a comprehensive rethinking of the communication system design, it is important to understand the fundamental limits on communication performance imposed by low-precision ADC. In this paper, we take a first step in this direction, investigating the Shannon-theoretic performance limits for the following idealized model: linear modulation over a real baseband Additive White Gaussian Noise (AWGN) channel with symbol rate Nyquist samples quantized by a low-precision ADC. This induces a discrete-time memoryless *AWGN-Quantized Output* channel, which is depicted in Figure 1. Under an *average* power constraint on the input power, we obtain the following results

1) For a $K$-level (i.e., $\log_2 K$ bits) output quantizer, we prove that the input distribution need not have any more than $K + 1$ mass points to achieve the channel capacity. (Numerical computation of optimal input distributions reveals that $K$ mass points are sufficient.) An intermediate result of interest is that, when the AWGN channel output is quantized with finite-precision, an average power constraint leads to an implicit peak power constraint, in the sense that an optimal input distribution must now have bounded support.

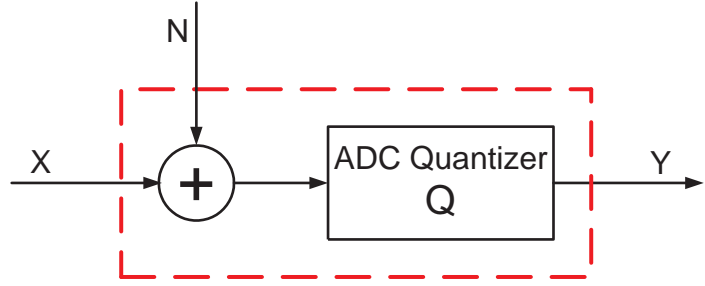

Fig. 1.   $Y = \mathsf{Q}(X + N)$ : The *AWGN-Quantized Ouput* channel induced by the output quantizer $\mathsf{Q}$.

2)  For 1-bit symmetric quantization, the preceding result can be tightened to show that binary antipodal signaling is optimal for any signal-to-noise ratio (SNR).

3)  For multi-bit quantizers, tight upper bounds on capacity are obtained using a dual formulation of the capacity problem. Near-optimal input distributions that approach these bounds are computed using the cutting-plane algorithm [31].

4)  While the preceding results optimize the input distribution for a fixed quantizer, comparison with an unquantized system requires an optimization over the choice of the quantizer as well. We numerically obtain optimal 2-bit and 3-bit symmetric quantizers.

5)  From our numerical results, we infer that low-precision ADC incurs a relatively small loss in spectral efficiency compared to unquantized observations. For example, 2-bit ADC achieves 95% of the spectral efficiency attained with unquantized observations at 0 dB SNR. Even at a moderately high SNR of 20 dB, 3-bit ADC achieves 85% of the spectral efficiency attained with unquantized observations. This indicates that DSP-centric design based on low-precision ADC is indeed attractive as communication bandwidths scale up, since the small loss in spectral efficiency should be acceptable in this regime. Furthermore, we also observe that a "sensible" choice of standard equiprobable pulse amplitude modulated (PAM) input with ADC thresholds set to implement maximum likelihood (ML) hard decisions achieves performance which is quite close to that obtained by numerical optimization of the quantizer and input distribution.

*Related Work*

For a Discrete Memoryless Channel (DMC), Gallager first showed that the number of input points with nonzero probability mass need not exceed the cardinality of the output [3, p. 96,

Corollary 3]. In our setting, the input alphabet is not a priori discrete, and there is a power constraint, so that the result in [3] does not apply. Our key result on the achievability of the capacity by a discrete input is actually an extension of a result of Witsenhausen in [4], where Dubins' theorem [5] was used to show that the capacity of a (discrete-time, memoryless and stationary) channel with $K$ output levels, under a *peak power* constraint is achievable by a discrete input with at most $K$ points. The key to our proof is to show that under output quantization, an average power constraint implies an implicit peak power constraint, after which we can use Dubins' theorem in a manner similar to the development in [4].

Prior work on the effect of reduced ADC precision on channel capacity with *fixed input distribution* includes [6], [7], [8]. However, other than our own preliminary results reported in [9], [10], we are not aware of a Shannon-theoretic investigation with low-precision ADC that includes optimization of the input distribution.

While we are interested in fundamental limits here, a strong motivation for this work comes from emergent applications in high-bandwidth, multiGigabit, unlicensed wireless communication systems using Ultrawideband (UWB) communication in the 3-10 GHz band [11], and millimeter wave communication in the 60 GHz band [12]. Indeed, there has been prior exploration of the impact of low-precision ADC in the specific context of UWB systems. Low power transceiver architectures for UWB systems have been proposed in [13], [14]. The performance of UWB receivers using 1-bit ADC has been analyzed in [15], including the use of dither and oversampling. The effect of ADC precision on UWB performance is considered in [16]. Decomposition of the UWB signal into parallel frequency channels in order to relax ADC speed requirements is considered in [17], [18]. Demodulation and interference suppression techniques for UWB communication using 1-bit ADC have been proposed in [19].

Given the encouraging results here, it becomes important to explore the impact of low-precision ADC on receiver tasks such as synchronization and equalization, which we have ignored in our idealized model (essentially assuming that these tasks have somehow already been accomplished). Related work on estimation using low-precision samples which may be relevant for this purpose includes the use of dither for signal reconstruction [20], [21], [22], frequency estimation using 1-bit ADC [23], [24], choice of quantization threshold for signal amplitude estimation [25], and signal parameter estimation using 1-bit dithered quantization [26], [27].

*Organization of the Paper*

The rest of the paper is organized as follows. The *AWGN-Quantized Output* channel model is described in the next section. In Section III, we show the existence of an implicit peak power constraint, and use it to prove that the capacity is achievable by a discrete input distribution. Section IV presents capacity computations, including duality-based upper bounds on capacity. Quantizer optimization is considered in Section V, followed by the conclusions in Section VI.

## II. CHANNEL MODEL

We consider linear modulation over a real AWGN channel, with symbol rate Nyquist samples quantized by a $K$-bin (or $K$-level) quantizer $\mathsf{Q}$. This induces the following discrete-time memoryless *AWGN-Quantized Output* (AWGN-QO) channel

$$Y = \mathsf{Q}\left(X + N\right) \ . \tag{1}$$

Here $X \in \mathbb{R}$ is the channel input with cumulative distribution function $F(x)$, $Y \in \{y_1, \cdots, y_K\}$ is the (discrete) channel output, and $N$ is $\mathcal{N}(0, \sigma^2)$ (the Gaussian random variable with mean $0$ and variance $\sigma^2$). $\mathsf{Q}$ maps the real valued input $X + N$ to one of the $K$ bins, producing a discrete output $Y$. In this work, we only consider quantizers for which each bin is an interval of the real line. The quantizer $\mathsf{Q}$ with $K$ bins is therefore characterized by the set of its $(K-1)$ thresholds $\boldsymbol{q} := [q_1, q_2, \cdots, q_{K-1}] \in \mathbb{R}^{K-1}$, such that $-\infty := q_0 < q_1 < q_2 < \cdots < q_{K-1} < q_K := \infty$. The output $Y$ is assigned the value $y_i$ when the quantizer input $(X + N)$ falls in the $i^{th}$ bin, which is given by the interval $(q_{i-1}, q_i]$. The resulting transition probability functions are

$$W_i(x) = \mathsf{P}(Y = y_i | X = x) = Q\left(\frac{q_{i-1} - x}{\sigma}\right) - Q\left(\frac{q_i - x}{\sigma}\right), \quad 1 \le i \le K, \tag{2}$$

where $Q(x)$ denotes the complementary Gaussian distribution function

$$Q(x) = \frac{1}{\sqrt{2\pi}} \int_x^\infty \exp(-t^2/2) dt \ .$$

The Probability Mass Function (PMF) of the output $Y$, corresponding to the input distribution $F$ is

$$R(y_i; F) = \int_{-\infty}^\infty W_i(x) dF(x), \quad 1 \le i \le K, \tag{3}$$

and the input-output mutual information $I(X; Y)$, expressed explicitly as a function of $F$ is

$$I(F) = \int_{-\infty}^{\infty} \sum_{i=1}^{K} W_i(x) \log \frac{W_i(x)}{R(y_i; F)} dF(x) .^{1} \tag{4}$$

Under an average power constraint $P$ on the channel input (i.e., $\mathbb{E}[X^2] \leq P$), we wish to compute the capacity of the channel (1), which is given by

$$C = \sup_{F \in \mathcal{F}} I(F), \tag{5}$$

where $\mathcal{F}$ is the set of all distributions on $\mathbb{R}$ that satisfy the average power constraint, i.e.,

$$\mathcal{F} = \left\{ F : \int_{-\infty}^{\infty} x^2 dF(x) \leq P \right\}. \tag{6}$$

## III. Discrete Input Distribution Achieves Capacity

We first employ the Karush-Kuhn-Tucker (KKT) optimality condition to show that, even though we have not imposed an explicit peak power constraint on the input, it is automatically induced by the average power constraint. Specifically, an optimal input distribution must have a bounded support set. This is then used to show that the capacity is achievable by a discrete input distribution with at most $K + 1$ mass points. Note that our result does not, however, guarantee that the capacity is achieved by a *unique* input distribution.

### A. An Implicit Peak Power Constraint

Using convex optimization principles, the following necessary and sufficient KKT optimality condition can be derived for our problem, in a manner similar to the development in [29], [30]. An input distribution $F$ is optimal for (5) *if and only if* there exists a $\gamma \geq 0$ such that

$$\sum_{i=1}^{K} W_i(x) \log \frac{W_i(x)}{R(y_i; F)} + \gamma(P - x^2) \leq I(F) \tag{7}$$

for all $x$, with equality if $x$ is in the support of $F$ [2], where the transition probability functions $W_i(x)$, and the output PMF $R(y_i; F)$ are as specified in (2) and (3), respectively.

The first term on the left hand side of the KKT condition is the Kullback-Leibler divergence (or the relative entropy), $D(W(\cdot|x)||R(\cdot; F))$, between the transition and the output distributions.

---

[1] The logarithm is base 2 throughout the paper, so the mutual information is measured in bits.

[2] The support of $F$ (or the set of increase points of $F$) is the set $S_X(F) = \{x : F(x + \epsilon) - F(x - \epsilon) > 0, \ \forall \epsilon > 0\}$.

For convenience, let us denote it by $d(x; F)$. We first study the behavior of this function in the limit as $x \to \infty$.

*Lemma 1:* For the AWGN-QO channel (1) with input distribution $F$, the divergence function $d(x; F)$ satisfies the following properties

(a) $\lim_{x \to \infty} d(x; F) = -\log R(y_K; F)$.

(b) There exists a finite constant $A_0$ such that $\forall \ x > A_0, d(x; F) < -\log R(y_K; F)$.

*Proof:* We have

$$d(x; F) = \sum_{i=1}^{K} W_i(x) \log \frac{W_i(x)}{R(y_i; F)}$$
$$= \sum_{i=1}^{K} W_i(x) \log(W_i(x)) - \sum_{i=1}^{K} W_i(x) \log(R(y_i; F)) \ .$$

For any finite noise variance $\sigma^2$, as $x \to \infty$, the conditional PMF $W_i(x)$ tends to the unit mass at $i = K$. This observation, combined with the fact that the entropy of a finite alphabet random variable is a continuous function of its probability law, gives

$$\lim_{x \to \infty} d(x; F) = 0 - \log(R(y_K; F)) = -\log(R(y_K; F)) \ .$$

To prove part (b), we pick $A_0$ to be such that $W_i(A_0) < R(y_i; F)$ for $i = \{1, 2, ..., K-1\}$, and also that $W_K(A_0) > R(y_K; F)$. Such an $A_0$ always exists because for $x > q_{K-1}$, the transition probabilities $W_i(x) \to 0$ and are strictly monotone decreasing functions of $x$ for $i = \{1, ..., K-1\}$, while $W_K(x) \to 1$ and is a strictly monotone increasing function of $x$ (the strict monotonicity is easy to see by evaluating the derivatives of the transition probabilities). With such a choice of $A_0$, we get that for $x > A_0$,

$$d(x; F) = \sum_{i=1}^{K} W_i(x) \log \frac{W_i(x)}{R(y_i; F)}$$
$$< W_K(x) \log \frac{W_K(x)}{R(y_K; F)} < -\log(R(y_K; F)).$$

$\blacksquare$

Using Lemma 1, we now prove the main result of this subsection.

*Proposition 1:* For the average power constrained AWGN-QO channel (1), an optimal input distribution must have bounded support.

*Proof:* Let us assume that the input distribution $F^*$ achieves [3] the capacity in (5), i.e., $I(F^*) = C$. Let $\gamma^* \geq 0$ denote a corresponding optimal Lagrange parameter, so that the KKT condition is satisfied. In other words, with $\gamma = \gamma^*$, and, $F = F^*$, (7) must be satisfied with an equality at every point in the support of $F^*$. We exploit this necessary condition next to show that the support of $F^*$ is upper bounded. Specifically, we prove that there exists a finite constant $A_2^*$ such that it is not possible to attain equality in (7) for any $x > A_2^*$.

Using Lemma 1, we first let $\lim_{x \to \infty} d(x; F^*) = -\log(R(y_K; F^*)) = L$, and also assume that there exists a finite constant $A_0$ such that $\forall \; x > A_0, d(x; F^*) < L$.

We consider two possible cases.

- Case 1: $\gamma^* > 0$.

  If $C > L + \gamma^* P$, then pick $A_2^* = A_0$.

  Else pick $A_2^* \geq \max\{A_0, \sqrt{(L + \gamma^* P - C)/\gamma^*}\}$.

  In either situation, for $x > A_2^*$, we get $d(x; F^*) < L$, and, $\gamma^* x^2 > L + \gamma^* P - C$.

  This gives

  $$d(x; F^*) + \gamma^*(P - x^2) < L + \gamma^* P - (L + \gamma^* P - C) = C.$$

- Case 2: $\gamma^* = 0$.

  Putting $\gamma^* = 0$ in the KKT condition (7), we get

  $$d(x; F^*) = \sum_{i=1}^{K} W_i(x) \log \frac{W_i(x)}{R(y_i; F^*)} \leq C \;, \quad \forall x.$$

  Thus,

  $$L = \lim_{x \to \infty} d(x; F^*) \leq C.$$

  Picking $A_2^* = A_0$, we therefore have that for $x > A_2^*$

  $$d(x; F^*) + \gamma^*(P - x^2) = d(x; F^*) < L.$$

  $$\implies d(x; F^*) + \gamma^*(P - x^2) < C.$$

Combining the two cases, we have shown that the support of the distribution $F^*$ has a finite upper bound $A_2^*$. Using similar arguments, it can easily be shown that the support of $F^*$ has a finite lower bound $A_1^*$ as well, which implies that $F^*$ has a bounded support. ∎

---

[3] That there exists an input which achieves the supremum in (5) is shown in Appendix I.

*B. Achievability of Capacity by a Discrete Input*

In [4], Witsenhausen considered a stationary discrete-time memoryless channel, with a continuous input $X$ taking values on the compact interval $[A_1, A_2] \subset \mathbb{R}$, and a discrete output $Y$ of finite cardinality $K$. It was shown that if the channel transition probability functions are continuous (i.e., $W_i(x)$ is continuous in $x$, for each $i = 1, \cdots, K$), then the capacity is achievable by a discrete input distribution with at most $K$ mass points. As stated in Theorem 1 below (proved in Appendix II), this result can be extended to show that, if an *additional* average power constraint is imposed on the input, the capacity is then achievable by a discrete input with at most $K + 1$ mass points.

*Theorem 1:* Consider a stationary discrete-time memoryless channel with a continuous input $X$ that takes values in the bounded interval $[A_1, A_2]$, and a discrete output $Y \in \{y_1, y_2, \cdots, y_K\}$. Let the channel transition probability function $W_i(x) = \mathsf{P}(Y = y_i | X = x)$ be continuous in $x$ for each $i$, where $1 \leq i \leq K$. The capacity of this channel, under an average power constraint on the input, is achievable by a discrete input distribution with at most $K + 1$ mass points.

*Proof:* See Appendix II. ∎

Theorem 1, coupled with the implicit peak power constraint derived in the previous subsection (Proposition 1), gives us the following result.

*Proposition 2:* The capacity of the average power constrained AWGN-QO channel (1) is achievable by a discrete input distribution with at most $K + 1$ points of support.

*Proof:* Using notation from the last subsection, let $F^*$ be an optimal distribution for (5), with the support of $F^*$ being contained in the bounded interval $[A_1{}^*, A_2{}^*]$. Define $\mathcal{F}_1$ to be the set of all average power constrained distributions $F$ whose support $S_X(F)$ is contained in $[A_1{}^*, A_2{}^*]$, i.e.,

$$\mathcal{F}_1 = \{F \in \mathcal{F} : S_X(F) \subseteq [A_1{}^*, A_2{}^*]\} \ , \tag{8}$$

where $\mathcal{F}$ is the set of all average power constrained distributions on $\mathbb{R}$, as defined in (6). Note that $F^* \in \mathcal{F}_1 \subset \mathcal{F}$. Consider the maximization of the mutual information $I(X; Y)$ over the set $\mathcal{F}_1$

$$C_1 = \max_{F \in \mathcal{F}_1} I(F). \tag{9}$$

Since the transition probability functions in (2) are continuous in $x$, Theorem 1 implies that a discrete distribution with at most $K + 1$ mass points achieves the maximum $C_1$ in (9). Denote such a distribution by $F_1$. However, since $F^*$ achieves the maximum $C$ in (5) and $F^* \in \mathcal{F}_1$, it must also achieve the maximum in (9). This implies that $C_1 = C$, and that $F_1$ is optimal for (5), thus completing the proof. ∎

### C. Symmetric Inputs for Symmetric Quantization

For our numerical capacity computations ahead, we assume that the quantizer Q employed in (1) is symmetric, i.e., its threshold vector $\boldsymbol{q}$ is symmetric about the origin. Given the symmetric nature of the AWGN noise and the power constraint, it seems intuitively plausible that restriction to symmetric quantizers should not be suboptimal from the point of view of optimizing over the quantizer choice in (1), although a proof of this conjecture has eluded us. However, once we assume that the quantizer in (1) is symmetric, we can restrict attention to only symmetric input distributions without loss of optimality, as stated in the following lemma.

*Lemma 2:* If the quantizer in (1) is symmetric, then, without loss of optimality, we can consider only symmetric input distributions (i.e., $F(x) = 1 - F(-x), \forall\ x \in \mathbb{R}$) for the capacity computation in (5).

*Proof:* Suppose we are given an input distribution $F(x)$ that is not necessarily symmetric. Consider now the following symmetric mixture distribution

$$\tilde{F}(x) = \frac{F(x) + 1 - F(-x)}{2}.$$

This mixture can be achieved by choosing distribution $F(x)$ or $1 - F(-x)$ with probability $1/2$ each. If we use $\tilde{F}(x)$ in place of $F(x)$, the conditional entropy $H(Y|X)$ remains unchanged due the symmetric nature of the noise $N$ and the quantizer. However, the output entropy $H(Y)$ changes as follows. Suppose that, when $F(x)$ is used, the PMF of $Y$ is $\boldsymbol{a} = [a_1, ..., a_M]$. Then under $1 - F(-x)$ it is $\hat{\boldsymbol{a}} = [a_M, ..., a_1]$. Hence under $\tilde{F}(x)$, the output $Y$ has the mixture PMF $\tilde{\boldsymbol{a}} = \frac{1}{2}(\boldsymbol{a} + \hat{\boldsymbol{a}})$. Since entropy is a concave function of the PMF,

$$H(Y)\big|_{Y \sim \tilde{\boldsymbol{a}}} \geq \frac{H(Y)\big|_{Y \sim \boldsymbol{a}}}{2} + \frac{H(Y)\big|_{Y \sim \hat{\boldsymbol{a}}}}{2} = H(Y)\big|_{Y \sim \boldsymbol{a}}.$$

It follows that under the symmetric distribution $\tilde{F}(x)$, $I(X;Y) = H(Y) - H(Y|X)$ is greater than that under $F(x)$, which proves the desired result. ∎

## IV. Capacity Computation

We now consider capacity computation for the AWGN-QO channel. We first provide an explicit capacity formula for the extreme scenario of 1-bit symmetric quantization, and then discuss numerical computations for multi-bit quantization.

### A. 1-bit Symmetric Quantization : Binary Antipodal Signaling is Optimal

With 1-bit symmetric quantization, the channel is

$$Y = \text{sign}(X + N). \tag{10}$$

Proposition 2 (section III-B) guarantees that the capacity of this channel is achievable by a discrete input distribution with at most 3 points. This result is further tightened by the following theorem that shows the optimality of binary antipodal signaling for all SNRs.

*Theorem 2:* For the 1-bit symmetric quantized channel model (10), the capacity is achieved by binary antipodal signaling and is given by

$$C = 1 - h\left(Q\left(\sqrt{\text{SNR}}\right)\right), \qquad \text{SNR} = \frac{P}{\sigma^2} \ ,$$

where $h(p)$ is the binary entropy function

$$h(p) = -p\log(p) - (1-p)\log(1-p) \ , \quad 0 \le p \le 1.$$

*Proof:* Since $Y$ is binary it is easy to see that

$$H(Y|X) = \mathbb{E}\left[h\left(Q\left(\frac{X}{\sigma}\right)\right)\right] \ ,$$

where $\mathbb{E}$ denotes the expectation operator. Therefore

$$I(X,Y) = H(Y) - \mathbb{E}\left[h\left(Q\left(\frac{X}{\sigma}\right)\right)\right] \ ,$$

which we wish to maximize over all input distributions satisfying $\mathbb{E}[X^2] \le P$. Since the quantizer is symmetric, we can restrict attention to symmetric input distributions without loss of optimality (cf. Lemma 2). On doing so, we obtain that the PMF of the output $Y$ is also symmetric (since the quantizer and the noise distribution are already symmetric). Therefore, $H(Y) = 1$ bit, and we obtain

$$C = 1 - \min_{\substack{X \text{ symmetric} \\ \mathbb{E}[X^2] \le P}} \mathbb{E}\left[h\left(Q\left(\frac{X}{\sigma}\right)\right)\right].$$

Since $h(Q(z))$ is an even function, we get that

$$H(Y|X) = \mathbb{E}\left[ h\left( Q\left( \frac{X}{\sigma} \right) \right) \right] = \mathbb{E}\left[ h\left( Q\left( \frac{|X|}{\sigma} \right) \right) \right].$$

In Appendix III, we show that the function $h(Q(\sqrt{y}))$ is convex in $y$. Thus, Jensen's inequality [32] implies that

$$H(Y|X) \geq h\left( Q\left( \sqrt{\mathsf{SNR}} \right) \right)$$

with equality iff $X^2 = P$. Coupled with the symmetry condition on $X$, this implies that binary antipodal signaling achieves capacity and the capacity is

$$C = 1 - h\left( Q\left( \sqrt{\mathsf{SNR}} \right) \right).$$

∎

### B. Multi-Bit Quantization

We now consider $K$-level quantization, where $K > 2$. It appears unlikely that closed form expressions for optimal input and capacity can be obtained, due to the complicated expression for mutual information. We therefore resort to the cutting-plane algorithm [31, Sec IV-A] to generate optimal inputs numerically. For channels with continuous input alphabets, the cutting-plane algorithm can, in general, be used to generate nearly optimal discrete input distributions. It is therefore well matched to our problem, for which we already know that the capacity is achievable by a discrete input distribution. It is worth mentioning that discretized Blahut-Arimoto type algorithms to compute the capacity of infinite input finite (infinite)-output channels have earlier been reported in [43], although they do not incorporate an average power constraint on the input.

We fix the noise variance $\sigma^2 = 1$, and vary the power $P$ to obtain capacity at different SNRs. To apply the cutting-plane algorithm, we take a fine quantized discrete grid on the interval $[-10\sqrt{P}, 10\sqrt{P}]$, and optimize the input distribution over this grid. Note that Proposition 1 (Section III-A) tells us that an optimal input distribution for our problem must have a bounded support, but it does not give explicit values that we can use directly in our simulations. However, on employing the cutting-plane algorithm over the interval $[-10\sqrt{P}, 10\sqrt{P}]$, we find that the resulting input distributions have support sets well within this interval. Moreover, increasing the interval length further does not change these results.

The input distributions generated by the cutting-plane algorithm are shown in our numerical results. We find that these distributions have support set cardinality less than $K + 1$ as predicted by Proposition 2. The optimality of these distributions can further be verified by comparing the mutual information they achieve with easily computable tight upper bounds on the capacity. The computation of these upper bounds is discussed next.

*1) Duality-Based Upper Bound on Channel Capacity:* In the dual formulation of the channel capacity problem, we focus on the distribution of the channel output, rather than that of the input. Specifically, assume a channel with input alphabet $\mathcal{X}$, transition law $W(y|x)$, and an average power constraint $P$. Then, for every choice of the output distribution $R(y)$, we have the following upper bound on the channel capacity $C$

$$C \leq U(R) = \min_{\gamma \geq 0} \sup_{x \in \mathcal{X}} [D(W(\cdot|x)||R(\cdot)) + \gamma(P - x^2)] \ , \tag{11}$$

where $\gamma$ is a Lagrange parameter, and $D(W(\cdot|x)||R(\cdot))$ is the divergence between the transition and output distributions. While [33] provides this bound for a Discrete Memoryless Channel (DMC), its extension to continuous alphabet channels has been established in [34], [35]. A detailed perspective on the use of duality-based upper bounds can be found in [36].

For an arbitrary choice of $R(y)$, the bound (11) might be quite loose. Therefore, to obtain a tight upper bound, we may need to evaluate (11) for a large number of output distributions and pick the minimum of the resulting upper bounds. This could be tedious in general, especially if the output alphabet is continuous. However, for the channel model we consider, the output alphabet is discrete with small cardinality. For example, for 2-bit quantization, the space of all output distributions is characterized by a set of just 3 parameters in the interval $(0, 1)$. This makes the dual formulation attractive, since we can easily obtain a tight upper bound on capacity by evaluating the upper bound in (11) for different choices of these parameters.

Next, we discuss computation of the upper bound (11) for our problem, for a fixed output distribution $R(y)$.

*Computation of the Upper Bound*: For convenience, we denote $d(x) = D(W(\cdot|x)||R(\cdot))$, and $g(x, \gamma) = d(x) + \gamma(P - x^2)$, so that we need to compute $\min_{\gamma \geq 0} \sup_{x \in \mathcal{X}} g(x, \gamma)$. Consider first the maximization over $x$, for a fixed $\gamma$. Although the input alphabet $\mathcal{X}$ is the real line $\mathbb{R}$, from a practical standpoint, we can restrict attention to a bounded interval $[M_1, M_2]$ while performing this maximization This is justified as follows. From Lemma 1, we know that $\lim_{x \to \infty} d(x) = \log \frac{1}{R(y_K)}$.

The saturating nature of $d(x)$, coupled with the non-increasing nature of $\gamma(P - x^2)$, implies that for all practical purposes, the search for the supremum of $d(x) + \gamma(P - x^2)$ over $x$ can be restricted to $x \leq M_2$, where $M_2$ is large enough to ensure that the difference $|d(x) - \log \frac{1}{R(y_K)}|$ is negligible for $x > M_2$. In our simulations, we take $M_2 = q_{K-1} + 5\sigma$, where $q_{K-1}$ is the largest quantizer threshold, and $\sigma^2$ is the noise variance. This choice of $M_2$ ensures that for $x > M_2$, the conditional PMF $W_i(x)$ is nearly the same as the unit mass at $i = K$, which consequently makes the difference between $d(x)$ and $\log \frac{1}{R(y_K)}$ negligible for $x > M_2$, as desired. Similarly, the search for the supremum over $x$ can also be restricted to $x \geq M_1 = q_1 - 5\sigma$, where $q_1$ is the smallest quantizer threshold. Note that if the quantizer and the output distribution $R(y)$ are picked to be symmetric, then the function $g(x, \gamma)$ is also symmetric in $x$, so that we can further restrict attention to $[0, M_2]$.

We now need to compute $\min_{\gamma \geq 0} \max_{x \in [M_1, M_2]} \{g(x, \gamma)\}$. To do this, we quantize the interval $[M_1, M_2]$ to generate a fine grid $\{x_1, x_2, \cdots, x_I\}$, and approximate the maximization over $x \in [M_1, M_2]$ as a maximization over this quantized grid. This reduces the computation of the upper bound to computing the function $\min_{\gamma \geq 0} \max_{1 \leq i \leq I} g(x_i, \gamma)$. Denoting $r_i(\gamma) := g(x_i, \gamma)$, this becomes $\min_{\gamma \geq 0} \max_{1 \leq i \leq I} r_i(\gamma)$. Hence, we are left with the task of minimizing (over $\gamma$) the maximum value of a finite set of functions of $\gamma$, which in turn can be done directly using the standard Matlab tool $fminimax$. Moreover, we note that the function being minimized over $\gamma$, i.e. $m(\gamma) := \max_{1 \leq i \leq I} r_i(\gamma)$, is convex in $\gamma$. This follows from the observation that each of the functions $r_i(\gamma) = d(x_i) + \gamma(P - x_i^2)$ is convex in $\gamma$ (in fact, affine in $\gamma$), so that their pointwise maximum is also convex in $\gamma$ [37, pp. 81]. The convexity of $m(\gamma)$ guarantees that $fminimax$ provides us the global minimum over $\gamma$.

*2) Numerical Results:* We now compare numerical results obtained using the cutting-plane algorithm with capacity upper bounds obtained using the preceding dual formulation. We fix the choice of quantizer to 2-bit symmetric quantization, in which case the quantizer is characterized by a single parameter $q$, with the quantizer thresholds being $\{-q, 0, q\}$. The results depicted in this section are for the particular quantizer choice $q = 2$.

The input distributions generated by the cutting-plane algorithm at various SNRs (setting $\sigma^2 = 1$) are shown in Figure 2, and the mutual information achieved by them is given in Table I. As predicted by Proposition 2 (section III-B), the support set of the input distribution (at each SNR) has cardinality $\leq 5$.

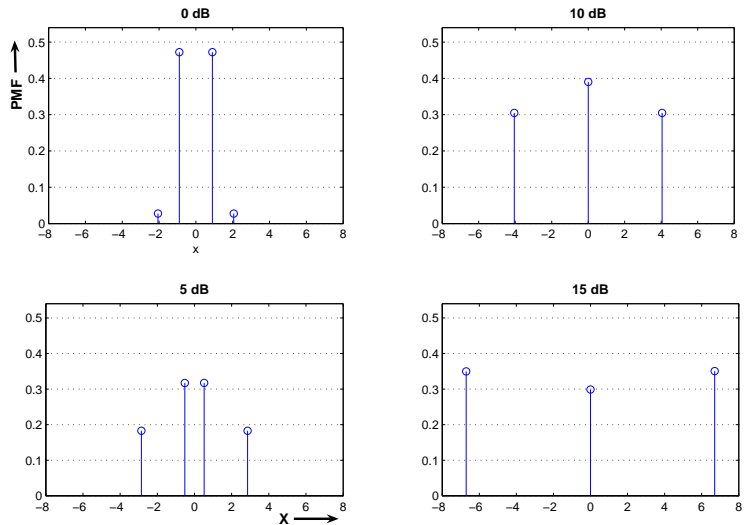

Fig. 2. Probability Mass Function of the optimal input generated by the cutting-plane algorithm at various SNRs, for the 2-bit symmetric quantizer with thresholds $\{-2, 0, 2\}$.

| SNR($dB$) | $-5$ | $0$ | $5$ | $10$ | $15$ | $20$ |
|-----------|------|-----|-----|------|------|------|
| Upper Bound | 0.1631 | 0.4055 | 0.8669 | 1.3859 | 1.5127 | 1.5146 |
| $MI$ | 0.1547 | 0.4046 | 0.8668 | 1.3792 | 1.4838 | 1.4839 |

TABLE I

DUALITY-BASED UPPER BOUNDS ON CHANNEL CAPACITY COMPARED WITH THE MUTUAL INFORMATION (MI) ACHIEVED BY THE DISTRIBUTIONS GENERATED USING THE CUTTING-PLANE ALGORITHM.

For upper bound computations, we evaluate (11) for different symmetric output distributions. For 2-bit quantization, the set of symmetric outputs is characterized by just one parameter $\alpha \in (0, 0.5)$, with the probability distribution on the output being $\{0.5 - \alpha, \alpha, \alpha, 0.5 - \alpha\}$. We vary $\alpha$ over a fine discrete grid on $(0, 0.5)$, and compute the upper bound for each value of $\alpha$. The least upper bound achieved thus, at a number of different SNRs, is shown in Table I

From the results, we see that the input distributions generated by the cutting-plane algorithm are nearly optimal, since they nearly achieve the capacity upper bound at every SNR. It is also insightful to look at the KKT condition for these input distributions. For instance, consider an SNR of 5 dB, for which the input distribution generated by the cutting-plane algorithm has support set $\{-2.86, -0.52, 0.52, 2.86\}$ and achieves a mutual information of 0.8668 bits. Figure 3 plots the function $g(x, \gamma)$ (i.e., the left hand side of the KKT condition (7)) for this input, with $\gamma = 0.1530$. We see that $g(x, \gamma)$ equals the mutual information at points in the support set of the input distribution, and is less than the mutual information everywhere else. The sufficient nature of the KKT condition therefore confirms the optimality of this input distribution. Note that we show the plot for $x \geq 0$ only because $g(x, \gamma)$ is symmetric in $x$.

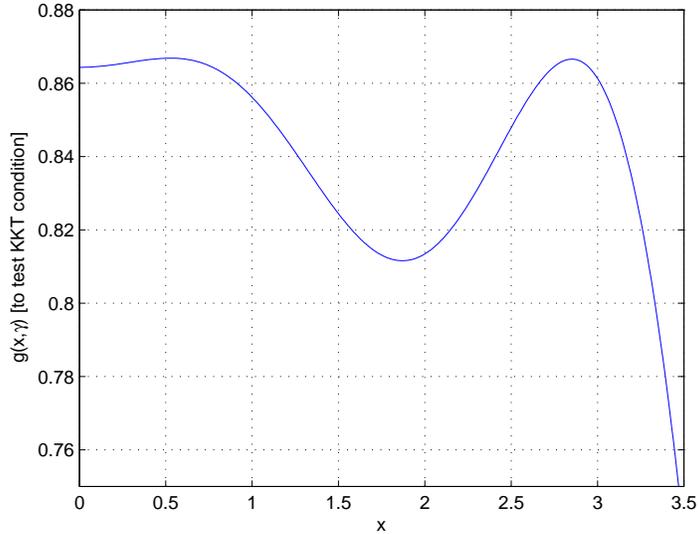

Fig. 3. KKT condition confirms the optimality of the input distribution generated by the cutting-plane algorithm.

## V. Optimization Over Quantizer

Till now, we have addressed the problem of capacity computation with a *fixed* output quantizer. The cutting-plane algorithm can be used to do this computation. In this section, we consider quantizer optimization, and numerically obtain optimal 2-bit and 3-bit symmetric quantizers.

*A Simple Benchmark:* While an optimal quantizer, along with a corresponding optimal input distribution, provides the absolute communication limits for our model, we do not have a simple analytical characterization of their dependence on SNR. From a system designer's perspective, therefore, it is of interest to also examine suboptimal choices that are easy to adapt as a function of SNR, as long as the penalty relative to the optimal solution is not excessive. Specifically, we take the following input and quantizer pair to be our *benchmark* strategy : for a $K$-level quantizer, consider equiprobable, equispaced $K$-PAM (Pulse Amplitude Modulation), with quantizer thresholds chosen to be the mid-points of the input mass point locations. That is, the quantizer levels correspond to the ML hard decision boundaries. Both the input mass points and the quantizer thresholds have a simple, well-defined dependence on SNR, and can therefore be adapted easily at the receiver based on the measured SNR. An explicit expression for the mutual information of our benchmark scheme is easy to compute. We can also obtain insight from the following lower bound on the mutual information, which is a direct consequence of Fano's inequality [32, pp. 37].

$$H_B(X|Y) \leq h(P_e) + P_e \log_2 (K-1) \ .$$

$$\implies I_B(X;Y) \geq \log_2 (K) - h(P_e) - P_e \log_2 (K-1) \ ,$$

where $h(\cdot)$ is the binary entropy function, and the subscript $B$ denotes the benchmark choice. The probability of error $P_e$ with the ML decisions is

$$P_e = 2 \left( \frac{K-1}{K} \right) Q \left( \sqrt{\frac{3 \ \mathsf{SNR}}{K^2-1}} \right) \ ,$$

where $Q(\cdot)$ is the complementary Gaussian distribution function.

It is evident that as $\mathsf{SNR} \to \infty$, $P_e \to 0$, so that $I_B(X;Y) \to \log_2(K)$ bits. This implies that the uniform PAM input with mid-point quantizer thresholds is near-optimal at high SNR. The issue to investigate therefore is how much gain an optimal quantizer and input pair provides over this benchmark at low to moderate SNR. Note that, for 1-bit symmetric quantization, the benchmark input corresponds to binary antipodal signalling, which has already been shown to be optimal for all SNRs.

As before, we set the noise variance $\sigma^2 = 1$ for convenience. Of course, the results are scale-invariant, in the sense that if both $P$ and $\sigma^2$ are scaled by the same factor $R$ (thus keeping the SNR unchanged), then there is an equivalent quantizer (obtained by scaling the thresholds by $\sqrt{R}$) that gives identical performance.

## NUMERICAL RESULTS

### A. 2-Bit Symmetric Quantization

A 2-bit symmetric quantizer is characterized by a single parameter $q$, with the quantizer thresholds being $\{-q, 0, q\}$. We therefore employ a brute force search over $q$ to find an optimal 2-bit symmetric quantizer. In Figure 4, we plot the variation of the channel capacity as a function of the parameter $q$ at various SNRs. Based on our simulations, we make the following observations

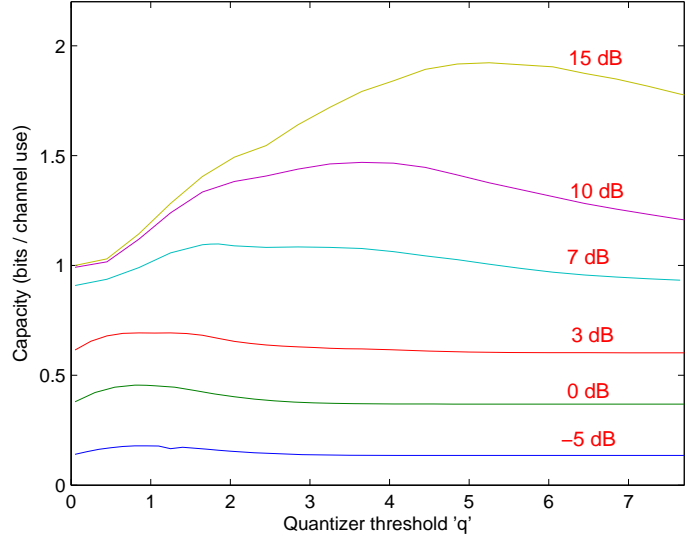

Fig. 4.   2-bit symmetric quantization : channel capacity (in bits per channel use) as a function of the quantizer threshold $q$ (noise variance assumed constant).

- For any SNR, there is an optimal choice of $q$ which maximizes capacity. For the benchmark quantizer (which is optimal at high SNR), $q$ scales as $\sqrt{\mathsf{SNR}}$, hence it is not surprising to note that the optimal value of $q$ we obtain increases monotonically with SNR at high SNR.

- The plots show that the capacity varies quite slowly as a function of $q$. This is because of the small variations in the channel transition probabilities (2) as a function of $q$.

- For any SNR, it is observed that, as $q \to 0$ or $q \to \infty$, we approach the same capacity as with 1-bit symmetric quantization (not shown for $q \to \infty$ in the plots for 10 and 15 dB in Figure 4). This conforms to intuition: $q = 0$ reduces the 2-bit quantizer to a 1-bit quantizer, while $q \to \infty$ renders the thresholds at $-q$ and $q$ ineffective in distinguishing between two finite valued inputs, so that only the comparison with the quantizer threshold at 0 yields useful information.

*Comparison with the Benchmark*: Table II compares the performance of the preceding optimal solutions with the benchmark scheme. The capacity with 1-bit symmetric quantization is also shown for reference. In addition to being nearly optimal at moderate to high SNRs, the benchmark scheme performs fairly well at low SNRs as well. For instance, even at -10 dB SNR, which might correspond to a UWB system designed for very low bandwidth efficiency, it achieves 86% of the capacity achieved with optimal choice of 2-bit quantizer and input distribution. On the other

| SNR(dB)        | $-20$  | $-10$  | $-5$   | 0      | 3      | 7      | 10     | 15     |
|----------------|--------|--------|--------|--------|--------|--------|--------|--------|
| 1-bit optimal  | 0.0046 | 0.0449 | 0.1353 | 0.3689 | 0.6026 | 0.9020 | 0.9908 | 0.9974 |
| 2-bit optimal  | 0.0063 | 0.0613 | 0.1792 | 0.4552 | 0.6932 | 1.0981 | 1.4731 | 1.9304 |
| 2-bit benchmark| 0.0049 | 0.0527 | 0.1658 | 0.4401 | 0.6868 | 1.0639 | 1.4086 | 1.9211 |

TABLE II

2-BIT SYMMETRIC QUANTIZATION : MUTUAL INFORMATION (IN BITS PER CHANNEL USE) ACHIEVED BY THE BENCHMARK

SCHEME, COMPARED AGAINST THE OPTIMAL SOLUTIONS.

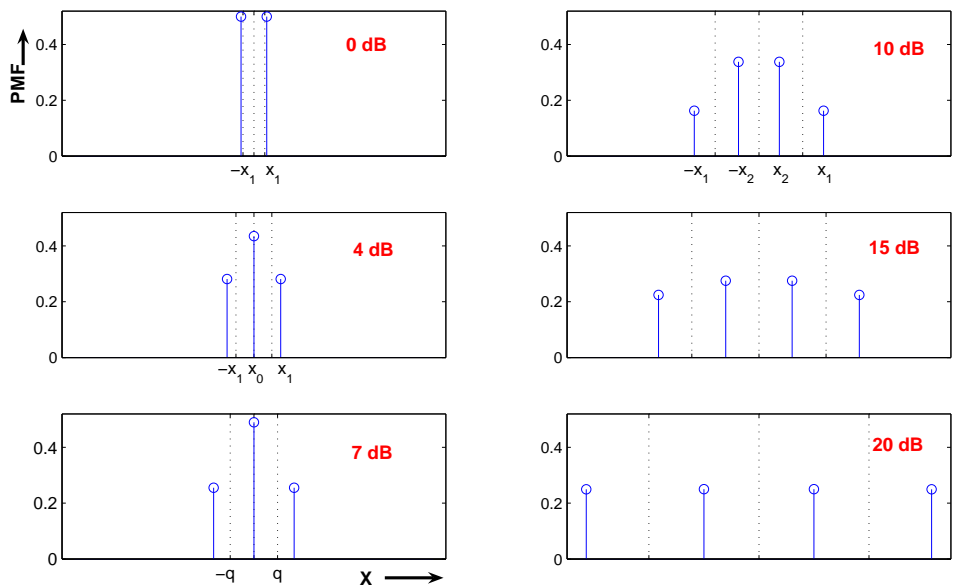

Fig. 5. 2-bit symmetric quantization : optimal input distribution and quantizer at various SNRs (the dashed vertical lines depict the locations of the quantizer thresholds).

hand, for SNR of 0 dB or above, the capacity is better than 95% of the optimal. These results are encouraging from a practical standpoint, given the ease of implementing the benchmark scheme.

*Optimal Input Distributions*: It is interesting to examine the optimal input distributions (given by the cutting-plane algorithm) corresponding to the optimal quantizers obtained above. Figure 5 shows these distributions, along with optimal quantizer thresholds, for different SNRs. The solid vertical lines show the locations of the input distribution points and their probabilities, while the quantizer thresholds are depicted by the dashed vertical lines. As expected, binary signaling is found to be optimal for low SNR, since it would be difficult for the receiver to distinguish between multiple input points located close to each other. The locations of the constellation points for the binary input are denoted by $\{-x_1, x_1\}$ in the $0$ dB plot in Figure 5. The number of mass points increases as SNR is increased, with a new point (denoted $x_0$) emerging at $0$. On increasing SNR further, we see that the points $\{-x_1, x_1\}$ (and also the quantizer thresholds $\{-q, q\}$) move farther apart, resulting in increased capacity. Finally, when the SNR becomes enough that four input points can be disambiguated, the point at $0$ disappears, and we get two new points shown at $\{-x_2, x_2\}$. The eventual convergence of this 4-point constellation to uniform PAM with mid-point quantizer thresholds (i.e., the benchmark scheme) is to be expected, since the benchmark scheme approaches the capacity bound of two bits at high SNR. It is worth noting that the optimal inputs we obtained all have at most four points, even though Proposition 2 (section III-B) is looser, guaranteeing the achievability of capacity by at most five points.

### B. 3-bit Symmetric Quantization

For 3-bit symmetric quantization, we need to optimize over a space of 3 parameters : $\{0 < q_1 < q_2 < q_3\}$, with the quantizer thresholds being $\{\pm q_1, \pm q_2, \pm q_3\}$. Since brute force search is computationally complex, we investigate an alternate iterative optimization procedure for joint optimization of the input and the quantizer in this case. Specifically, we begin with an initial quantizer choice $Q_1$, and then iterate as follows (starting at $i = 1$)

- For the quantizer $Q_i$, find an optimal input. Call this input $F_i$.
- For the input $F_i$, find a locally optimal quantizer, initializing the search at $Q_i$. Call the resulting quantizer $Q_{i+1}$.
- Repeat the first two steps with $i = i + 1$.

We terminate the process when the capacity gain between consecutive iterations becomes less than a small threshold $\epsilon$.

Although the input-output mutual information is a concave functional of the input distribution (for a fixed quantizer), it is not guaranteed to be concave jointly over the input and the quantizer. Hence, the iterative procedure is not guaranteed to provide an optimal input-quantizer pair in general. A good choice of the initial quantizer $Q_1$ is crucial to enhance the likelihood that it does converge to an optimal solution. We discuss this next.

| SNR(dB) | $-20$ | $-10$ | $-5$ | 0 | 5 | 10 | 15 | 20 |
|---------|-------|-------|------|---|---|----|----|----|
| 3-bit optimal | 0.0069 | 0.0667 | 0.1926 | 0.4817 | 0.9753 | 1.5844 | 2.2538 | 2.8367 |
| 3-bit benchmark | 0.0050 | 0.0557 | 0.1768 | 0.4707 | 0.9547 | 1.5332 | 2.1384 | 2.8084 |

TABLE III

3-BIT SYMMETRIC QUANTIZATION : MUTUAL INFORMATION (IN BITS PER CHANNEL USE) ACHIEVED BY THE BENCHMARK SCHEME, COMPARED AGAINST THE OPTIMAL SOLUTIONS.

*High SNR Regime*: For high SNRs, we know that the uniform PAM with mid-point quantizer thresholds (i.e., the benchmark scheme) is nearly optimal. Hence, this quantizer is a good choice for initialization at high SNRs. The results we obtain indeed demonstrate that this initialization works well at high SNRs. This is seen by comparing the results of the iterative procedure with the results of a brute force search over the quantizer choice (similar to the 2-bit case considered earlier), as both of them provide almost identical capacity values.

*Lower SNRs*: For lower SNRs, one possibility is to try out different initializations $Q_1$. However, on trying out the benchmark initialization at some lower SNRs as well, we find that the iterative procedure still provides us with near optimal solutions (again verified by comparing with brute force optimization results). While our results show that the iterative procedure (with benchmark initialization) has provided (near) optimal solutions at different SNRs, we leave the question of whether it will converge in general to an optimal solution or not as an open problem.

*Comparison with the Benchmark*: The efficacy of the benchmark initialization at lower SNRs suggests that the performance of the benchmark scheme should not be too far from optimal at small SNRs as well. This is indeed the case, as shown in Table III. At 0 dB SNR, for instance, the benchmark scheme achieves $98\%$ of the capacity achievable with an optimal quantizer choice.

*Optimal Input Distributions*: The optimal input distributions and quantizers (obtained using the iterative procedure) are depicted in Figure 6. Binary antipodal signaling is optimal at low SNRs (not shown). Increase in the SNR first results in a new mass point at 0, and subsequently in a 4-point constellation. The trend is repeated, with the number of mass points increasing with SNR, till we get an 8-point constellation which eventually moves towards uniform PAM, and the capacity approaches three bits.

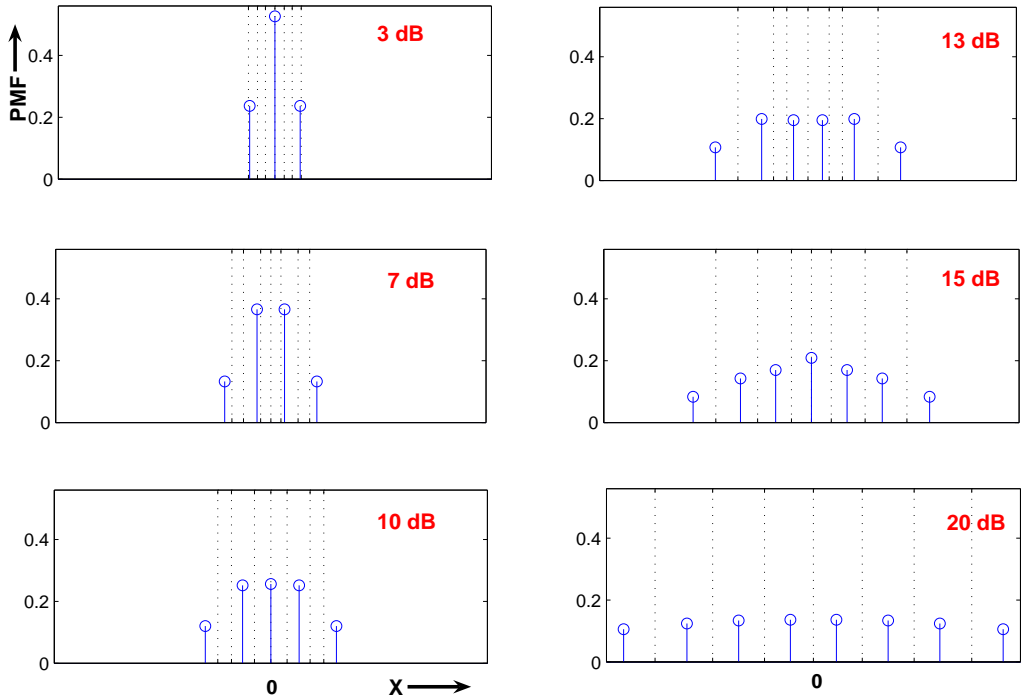

Fig. 6. 3-bit symmetric quantization : optimal input distribution and quantizer at various SNRs (the dashed vertical lines depict the locations of the quantizer thresholds).

Again, the optimal input distributions obtained have at most $K$ points ($K = 8$), while Proposition 2 in section III-B provides the looser guarantee that the capacity is achievable with at most $K+1$ points. Of course, the results above are for the particular cases when the quantizers are also optimal (among symmetric quantizers), whereas Proposition 2 holds for any quantizer choice. Thus, it is possible that there might exist a $K$-level quantizer for which the capacity is indeed achieved by exactly $K+1$ points. We leave open, therefore, the question of whether the result in Proposition 2 can be tightened to guarantee the achievability of capacity with at most $K$ points.

## C. Comparison with Unquantized Observations

We now compare the capacity results for different quantizer precisions against the capacity with unquantized observations (depicted in Figure 7). A sampling of these results is provided in Table IV. We observe that at low SNR, the performance degradation due to low-precision quantization is small. For instance, at -5 dB SNR, 1-bit receiver quantization achieves 68% of the capacity achievable with infinite-precision, while with 2-bit quantization, we can get as much as 90% of

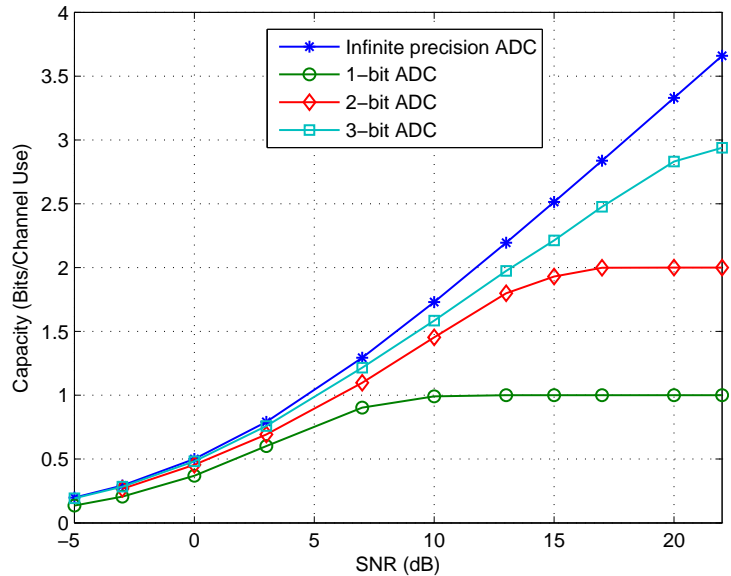

Fig. 7. Capacity with 1-bit, 2-bit, 3-bit, and infinite-precision ADC.

| SNR(dB) | −10 | −5 | 0 | 5 | 10 | 15 | 20 |
|---|---|---|---|---|---|---|---|
| 1-bit ADC | 0.0449 | 0.1353 | 0.3689 | 0.7684 | 0.9908 | 0.9999 | 0.9999 |
| 2-bit ADC | 0.0613 | 0.1792 | 0.4552 | 0.8889 | 1.4731 | 1.9304 | 1.9997 |
| 3-bit ADC | 0.0667 | 0.1926 | 0.4817 | 0.9753 | 1.5844 | 2.2538 | 2.8367 |
| Unquantized | 0.0688 | 0.1982 | 0.5000 | 1.0286 | 1.7297 | 2.5138 | 3.3291 |

TABLE IV

IMPACT OF LOW-PRECISION ADC : CAPACITY (IN BITS PER CHANNEL USE) WITH DIFFERENT ADC PRECISIONS, COMPARED WITH THE UNQUANTIZED (INFINITE-PRECISION) CASE.

the infinite-precision limit. This is to be expected: if channel noise dominates the actual signal, increasing the quantizer precision beyond a point does not help much in distinguishing between different signal levels. The more surprising finding is that, even at moderately high SNRs, the loss due to low-precision sampling remains quite acceptable. For example, 2-bit quantization achieves 85% of the capacity attained using unquantized observations at 10 dB SNR, while 3-bit quantization achieves 85% of the unquantized capacity at 20 dB SNR. Encouraging results of a similar nature have been reported earlier in [6]. However, the input alphabet there was kept fixed as binary to begin with, so that the good performance with low-precision receiver quantization is perhaps less surprising.

|  | Spectral Efficiency (bits per channel use) | | | | |
|---|---|---|---|---|---|
|  | 0.25 | 0.5 | 1.0 | 1.73 | 2.5 |
| 1-bit ADC | $-2.04$ | 1.79 | — | — | — |
| 2-bit ADC | $-3.32$ | 0.59 | 6.13 | 12.30 | — |
| 3-bit ADC | $-3.67$ | 0.23 | 5.19 | 11.04 | 16.90 |
| Unquantized | $-3.83$ | 0.00 | 4.77 | 10.00 | 14.91 |

TABLE V

SNR (IN DB) REQUIRED TO ACHIEVE A SPECIFIED SPECTRAL EFFICIENCY WITH DIFFERENT ADC PRECISIONS.

While the loss in spectral efficiency at fixed SNR is moderate, the loss in power efficiency at fixed spectral efficiency is significant (Table V). For example, if the spectral efficiency is fixed to that attained by an unquantized system at 10 dB (which is 1.73 bits/channel use), then 2-bit quantization incurs a loss of 2.30 dB. In practical terms, this penalty in power is more significant compared to the 15% loss in spectral efficiency on using 2-bit quantization at 10 dB SNR. This suggests, for example, that in order to weather the impact of low-precision ADC, a moderate reduction in the spectral efficiency is a better design choice than an increase in the transmit power.

## VI. CONCLUSIONS

Our Shannon-theoretic investigation indicates that the use of low-precision ADC is a feasible option for designing future high-bandwidth communication systems. The choice of low-precision ADC is consistent with the overall system design goals for systems such as UWB and mm wave communication, where power is at a premium, due to regulatory restrictions as well as due to the difficulty of generating large transmit powers with integrated circuits in low-cost silicon processes (e.g., see [38] for discussion of mm wave CMOS design). Power-efficient communication dictates the use of small constellations, so that the symbol rate, and hence the sampling rate, for a given bit rate must be high. This forces us towards using ADCs with lower precision, but fortunately, this is consistent with the use of small constellations in the first place for power-efficient design. Thus, if we plan on operating at low to moderate SNR, the small reduction in spectral efficiency due to low-precision ADC is acceptable in such systems, given that bandwidth is plentiful.

There are several unresolved technical issues that we leave as open problems. While we show that at most $K + 1$ points are needed to achieve capacity for a $K$-level quantizer, our numerical results show that at most $K$ points are needed. Can this be proven, at least for symmetric quantizers? Are symmetric quantizers optimal? Does our iterative procedure (with the benchmark initialization, or some other judicious initialization) for joint optimization of the input and the quantizer converge to an optimal solution in general? Are there other, provably optimal techniques with substantially lower complexity than brute force search to perform this joint optimization?

A technical assumption worth revisiting is that of Nyquist sampling (which induces the discrete-time memoryless AWGN-Quantized Output channel model considered in this work). While symbol rate Nyquist sampling is optimal for unquantized systems in which the transmit and receive filters are square root Nyquist and the channel is ideal, for quantized samples, we have obtained numerical results that show that fractionally spaced samples can actually lead to small performance gains. A detailed study quantifying such gains is important in understanding the tradeoffs between ADC speed and precision. However, we do not expect oversampling to play a significant role at low to moderate SNR, given the small degradation in our Nyquist sampled system relative to unquantized observations (for which Nyquist sampling is indeed optimal) in these regimes. Of course, oversampling in conjunction with hybrid analog/digital processing (e.g., using ideas analogous to delta-sigma quantization) could produce bigger performance gains, but this falls outside the scope of the present model.

While our focus in this paper was on non-spread systems, it is known that low-precision ADC is often employed in spread spectrum systems for low cost implementations [39]. In our prior examination of Shannon limits for direct sequence spread spectrum systems with 1-bit ADC [9], we demonstrated that binary signaling was suboptimal, but did not provide a complete characterization of an optimal input distribution. The approach in the present paper implies that, for a spreading gain $G$, a discrete input distribution with at most $G + 2$ points can achieve capacity (although in practice, much smaller constellations would probably work well).

Finally, we would like to emphasize that the Shannon-theoretic perspective provided in this paper is but a first step towards the design of communication systems with low-precision ADC. Major technical challenges include the design of ADC-constrained methods for carrier and timing synchronization, channel equalization, demodulation and decoding.

## Appendix I : Achievability of Capacity

*Theorem 3:* [40], [41] Let $\mathcal{V}$ be a real normed linear vector space, and $\mathcal{V}^*$ be its normed dual space.

(a) A weak* continuous real-valued functional $f$ evaluated on a weak* compact subset $\mathcal{F}$ of $\mathcal{V}^*$ achieves its maximum on $\mathcal{F}$.

(b) If in addition to part (a), $\mathcal{F}$ is a convex subset, and $f$ is a convex functional, then the maximum is achieved at an extreme point [4] of $\mathcal{F}$.

*Proof:* For part (a), see [40, p. 128, Thm 2]. Part (b) follows from the Bauer Maximum Principle (see, for example [41, p. 211]), which holds since the dual space $\mathcal{V}^*$, equipped with the weak* topology, is a locally convex Hausdorff space [41, p. 205]. ∎

The use of part (a) of the theorem to establish the existence of capacity-achieving input distributions is standard (see [30], [42] for details). To use this theorem for our channel model (1), we need to show that the set $\mathcal{F}$ of all average power constrained distribution functions is weak* compact, and the mutual information functional $I$ is weak* continuous over $\mathcal{F}$, so that $I$ achieves its maximum on $\mathcal{F}$. The weak* compactness of $\mathcal{F}$ follows by [42, Lemma 3.1]. To prove continuity, we need to show that

$$F_n \xrightarrow{weak^*} F \implies I(F_n) \longrightarrow I(F)$$

The finite cardinality of the output for our problem trivially ensures this. Specifically,

$$I(F) = H_Y(F) - H_{Y|X}(F)$$
$$= -\sum_{i=1}^{K} R(y_i; F) \log R(y_i; F) + \int dF(x) \sum_{i=1}^{K} W_i(x) \log W_i(x)$$

where,

$$R(y_i; F) = \int_{-\infty}^{\infty} W_i(x) dF(x).$$

The continuous and bounded nature of $W_i(x)$ ensures that $R(y_i; F)$ is continuous (by the definition of weak* topology). Moreover, the function $\sum_{i=1}^{K} W_i(x) \log W_i(x)$ is also continuous and bounded, implying that $H_{Y|X}(F)$ is also continuous (again by the definition of weak* topology). The continuity of $I(F)$ thus follows.

---

[4]An extreme point of a convex set $\mathcal{F}$ is a point that is not obtainable as a mid-point of two distinct points of $\mathcal{F}$.

## Appendix II : Proof of Theorem 1 (Discrete Capacity-Achieving Distribution)

The proof is along the same lines as Witsenhausen's proof in [4], except that we have an additional average power constraint on the input.

*Proof:* Let $\mathcal{S}$ be the set of all average power constrained distributions with support in the interval $[A_1, A_2]$. The required capacity, by definition, is $C = \sup_{\mathcal{S}} I(X; Y)$, where $I(X; Y)$ denotes the mutual information between $X$ and $Y$. The achievability of the capacity is guaranteed by Theorem 3(a) in Appendix I. [42, Lemma 3.1] ensures the weak* compactness of the set $S$, while weak* continuity of $I(X; Y)$ is easily proven given the assumption that the transition functions $W_i(x)$ are continuous. Let $S^*$ be a capacity-achieving input distribution.

The key idea that we employ is a theorem by Dubins [5], which characterizes extreme points of the intersection of a convex set with hyperplanes. We first give some necessary definitions, and then state the theorem.

*Definitions :*

- Let $\mathcal{E}$ be a vector space over the field of real numbers, and $\mathcal{M}$ be a convex subset of $\mathcal{E}$. $\mathcal{M}$ is said to be *linearly bounded* (*linearly closed*) if every line intersects $\mathcal{M}$ in a bounded (closed) subset of the line.

- Let $f : \mathcal{E} \to \mathbb{R}$ be a linear functional (not identically zero). The set $\{x \in \mathcal{E} : f(x) = c\}$ defines a hyperplane, for any real $c$.

*Dubins' Theorem :* Let $\mathcal{M}$ be a linearly closed and linearly bounded convex set and $\mathcal{U}$ be the intersection of $\mathcal{M}$ with $n$ hyperplanes, then every extreme point of $\mathcal{U}$ is a convex combination of at most $n + 1$ extreme points of $\mathcal{M}$.

To apply Dubins' theorem to our problem, we begin by defining $C[A_1, A_2]$ : the real normed linear space of all continuous functions on the interval $[A_1, A_2]$, with sup-norm. The dual of $C[A_1, A_2]$ is the space of functions of bounded variations [40, Sec 5.5], and it includes the (convex) set of all distribution functions with support in $[A_1, A_2]$. We take $\mathcal{E}$ to be the dual of $C[A_1, A_2]$, and $\mathcal{M}$ to be the subset of $\mathcal{E}$ consisting of all distribution functions with support in $[A_1, A_2]$. Note that the optimal input distribution $S^* \in \mathcal{M}$.

Let the probability vector of the output $Y$, when the input is $S^*$, be $R^* = \{p_1^*, p_2^*, \cdot, p_K^*\}$. Also, let the average power of the input under the distribution $S^*$ be $P_0$, where $P_0 \leq P$.

Now, consider the following subset $\mathcal{U}$ of $\mathcal{M}$

$$\mathcal{U} = \{M \in \mathcal{M} | R(y; M) = R^* \text{ and } E(X^2) = P_0\}. \tag{12}$$

The set $\mathcal{U}$ is the intersection of the set $\mathcal{M}$ with the following $K$ hyperplanes

$$H_i : \int_{A_1}^{A_2} W_i(x) dM(x) = p_i^* \quad 1 \leq i \leq K-1 \tag{13}$$

and,

$$H_K : \int_{A_1}^{A_2} x^2 dM(x) = P_0 \tag{14}$$

where $W_i(x)$ are the transition probability functions. Note that there are only $K-1$ hyperplanes in (13) since the probabilities must sum to 1, thus making the requirement on $p_K^*$ redundant.

We know that the set $\mathcal{M}$ is compact in the weak* topology [42, Lemma 3.1]. Also, each of the hyperplanes $H_i, 1 \leq i \leq K-1$, is a closed set since the functions $W_i(x)$ are continuous. The hyperplane $H_K$ is closed as well, since $x^2$ is a continuous function. Therefore, the set $\mathcal{U}$, being the intersection of a weak* compact set with $K$ closed sets, is weak* compact. It is easy to see that $\mathcal{U}$ is a convex set as well. On the set $\mathcal{U}$, we have

$$I(X; Y) = H(Y) - H(Y|X)$$

$$= -\sum_{i=1}^{K} p_i^* \log p_i^* + \int_{A_1}^{A_2} dM(x) \sum_{i=1}^{K} W_i(x) \log W_i(x).$$

As a function of the distribution $M(\cdot)$, we get

$$I(X; Y) = \text{ constant } + \text{ linear },$$

and the linear part is weak* continuous since $\sum_{i=1}^{K} W_i(x) \log W_i(x)$ is in $C[A_1, A_2]$.

It follows from Theorem 3(b) in Appendix I that the continuous linear functional $I(X; Y)$ attains its maximum over the compact convex set $\mathcal{U}$ at an extreme point of $\mathcal{U}$. However, since $S^* \in \mathcal{U}$, any maxima over $\mathcal{U}$ is a maxima over $\mathcal{S}$ as well. Hence, the required capacity is achieved at an extreme point of $\mathcal{U}$.

We now apply Dubins' theorem to characterize the extreme points of $\mathcal{U}$. Since $\mathcal{U}$ is the intersection of $\mathcal{M}$ with $K$ hyperplanes, every extreme point of $\mathcal{U}$ is a convex combination of at most $K+1$ extreme points of $\mathcal{M}$. The extreme points of $\mathcal{M}$ however are distributions concentrated at single points within the interval $[A_1, A_2]$. Therefore, we get that the required capacity is achievable by a discrete distribution with at most $K+1$ points of support. ∎

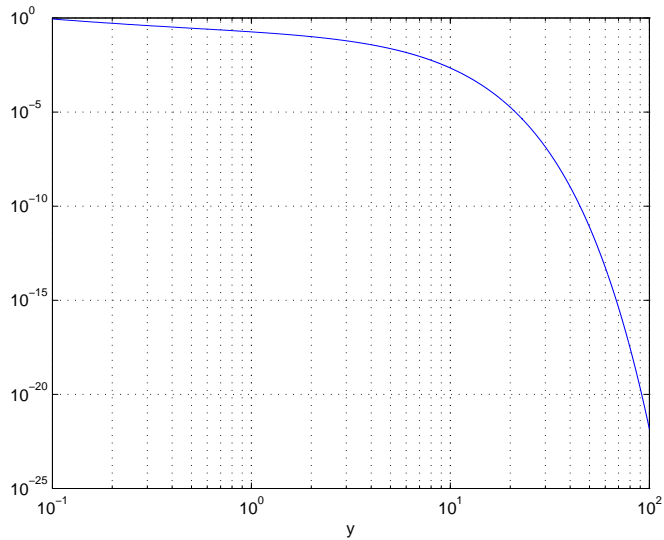

Fig. 8.   The second derivative of $h(Q(\sqrt{y}))$ is positive everywhere.

## Appendix III : Convexity of the Function $h(Q(\sqrt{y}))$

To show convexity, we verify that the second derivative of the function $h(Q(\sqrt{y}))$ is positive everywhere. For $y > 2$, we do this analytically, while for $0 \leq y \leq 2$, the positivity of the second derivative is demonstrated numerically in Figure 8.

Let $u(y) = h(Q(\sqrt{y}))$. Then,

$$u'(y) = \frac{-e^{-y/2}}{2\sqrt{2\pi y}\ln 2} \ln\left(\frac{1 - Q(\sqrt{y})}{Q(\sqrt{y})}\right)$$

Note that $\frac{1-Q(\sqrt{y})}{Q(\sqrt{y})} \geq 1, \forall y \geq 0$. Therefore, to show that the second derivative $u''(y)$ is positive, it suffices to show that the function $v(y) = e^{-y/2}\ln\left[\frac{1-Q(\sqrt{y})}{Q(\sqrt{y})}\right]$ is a decreasing function of $y$. Taking the derivative of $v(y)$, we get

$$v'(y) = \frac{-e^{-y/2}}{2}\left[\ln\left(\frac{1-Q(\sqrt{y})}{Q(\sqrt{y})}\right) - \frac{e^{-y/2}}{\sqrt{2\pi y}\,Q(\sqrt{y})(1-Q(\sqrt{y}))}\right]$$

To show that $v(y)$ is decreasing, it suffices to show that

$$\ln\left(\frac{1-Q(\sqrt{y})}{Q(\sqrt{y})}\right) \geq \frac{e^{-y/2}}{\sqrt{2\pi y}\,Q(\sqrt{y})(1-Q(\sqrt{y}))} \tag{15}$$

Using the fact [28, pp. 78] that $Q(y) \geq (1 - \frac{1}{y^2})\frac{e^{-y^2/2}}{y\sqrt{2\pi}}$, we get that if $y > 1$, then the following condition is sufficient for (15) to be true

$$\ln\left(\frac{1-Q(\sqrt{y})}{Q(\sqrt{y})}\right) \geq \frac{1}{(1-\frac{1}{y})(1-Q(\sqrt{y}))} \tag{16}$$

or, equivalently

$$(1 - \frac{1}{y})(1 - Q(\sqrt{y})) \ln \left( \frac{1 - Q(\sqrt{y})}{Q(\sqrt{y})} \right) \geq 1 \qquad (17)$$

The left hand side of (17) is a monotone increasing function of $y$. For $y = 2$, it equals $1.133$. Thus (17) holds $\forall y > 2$, and hence the second derivative of $h(Q(\sqrt{y}))$ must be positive for $y > 2$.

## ACKNOWLEDGMENT

The first author would like to thank Prof. Janos Englander (Department of Probability and Statistics, UCSB), Prof. Shiv Chandrasekaran (Department of Electrical and Computer Engineering, UCSB) and Mr. Vinay Melkote (Signal Compression Laboratory, UCSB) for several useful discussions.